\documentstyle[12pt,aaspp4]{article}
\begin{document}
\parindent=1.0cm

\title{Photometric Variability among the Brightest Asymptotic Giant Branch 
Stars Near the Center of M32\footnote[1]{Based on observations obtained at the
Gemini Observatory, which is operated by the Association of Universities
for Research in Astronomy, Inc., under a co-operative agreement with the
NSF on behalf of the Gemini partnership: the National Science Foundation
(United States), the Particle Physics and Astronomy Research Council
(United Kingdom), the National Research Council of Canada (Canada),
CONICYT (Chile), the Australian Research Council (Australia), CNPq (Brazil),
and CONICET (Argentina).}}

\author{T. J. Davidge}

\affil{Herzberg Institute of Astrophysics,
\\National Research Council of Canada, 5071 West Saanich Road,
\\Victoria, B.C. Canada V9E 2E7\\ {\it email: tim.davidge@nrc.ca}}

\author{F. Rigaut}

\affil{Gemini Observatory, 670 North A'ohoku Place,
\\Hilo, HI 96720-2700\\ {\it email: frigaut@gemini.edu}}

\begin{abstract}

	Deep $K'$ images with 0.1 arcsec angular resolution, 
obtained with ALTAIR$+$NIRI on Gemini North, are used to investigate 
photometric variablity among the brightest asymptotic giant branch (AGB) stars in 
the central regions of M32. Based on a comparison with brightnesses obtained from 
the $K-$band data discussed by Davidge et al. (2000, ApJ, 545, L89), it is 
concluded that (1) at least 60\% of bright AGB stars near the center of M32 are 
photometrically variable, and (2) the amplitudes of the light variations are 
similar to those of long period variables in the Galactic bulge. We do not find 
evidence for a population of large amplitude variables, like those detected by 
IRAS in the Galactic bulge. The technique discussed here may prove 
useful for conducting an initial reconnaisance of photometric variability among 
AGB stars in spheroids in the Virgo cluster and beyond, where the required long 
exposure times may restrict observations to only a few epochs.

\end{abstract}


\section{INTRODUCTION}

	Studies of nearby galaxies lay the 
groundwork for understanding the evolution of more distant systems. While lacking 
a massive classical elliptical galaxy, the Local Group contains the compact 
elliptical galaxy M32, which served as the basis for early pioneering efforts to 
understand the stellar contents of distant galaxies (e.g. O'Connell 1980; Rose 
1985). However, the nature of M32, and its relation to classical elliptical 
galaxies, remains a matter of on-going debate. While it has been 
suggested that interactions with M31 played a critical role in defining the 
present-day properties of the galaxy (e.g. Faber 1973, Burkert 1994; Bekki et al. 
2001), the central structural characteristics of M32
suggest that its gross properties have not changed 
drastically since it formed (Graham 2002; Choi, Guhathakurta, \& Johnston 2002). 

	Studies of the stellar content of M32 will provide insights into its 
evolution, and the brightest stars are convenient probes for this 
purpose. The brightest red stars in M32 are evolving on the 
asymptotic giant branch (AGB), and a number of studies have investigated the 
resolved stellar content in the outer (Freedman 1989, 1992; Elston \& Silva 1992; 
Davidge \& Jones 1992, Grillmair et al. 1996, Davidge 2000) and inner (Davidge 
et al. 2000) regions of this galaxy. The number density, normalized with respect 
to red surface brightness, and peak brightness of the brightest AGB 
stars do not change with radius (Davidge 2000; Davidge et al. 2000), 
indicating that these stars are well mixed throughout 
the main body of M32, and likely belong to a 
population that formed when its present-day morphology was 
imprinted. The bright AGB content of M32 also appears to be representative of 
the inner bulge of M31 and more distant systems (Davidge 2001, 2002).

	Many of the brightest AGB stars are expected to be long period variables 
(LPVs), and studies of these objects will provide clues into the stellar 
content of M32. For example, the detection of very large amplitude LPVs, 
such as those that were detected by IRAS in the Galactic bulge (e.g. Glass 
et al. 1995, 2001), would be one signature of an intermediate age population. 
In the present paper, high angular resolution $K-$band 
observations of the brightest AGB stars in the central regions of M32 
are used to investigate photometric variablity among these objects. 
A comparison is made with the LPV content of the Galactic bulge.

\section{OBSERVATIONS AND DATA REDUCTION}

	A series of 60 sec $K'$ exposures were recorded of the central regions of 
M32 on the night of UT October 12 2003 as part of the system verification 
program for ALTAIR$+$NIRI. ALTAIR is the facility adaptive optics (AO) 
system on Gemini North (GN), and a detailed description of this device has been 
given by Herriot et al. (2000). The nucleus of M32 
served as the reference beacon for AO compensation. 
NIRI, which is the facility infrared imager on GN, was 
used in f/32 mode for these observations, and so the image scale is 
0.022 arcsec pixel$^{-1}$; the $1024 \times 1024$ InSb array in NIRI 
thus images a $22 \times 22$ arcsec field.

	The data were reduced with a standard pipeline for infrared images (e.g. 
Davidge \& Courteau 1999). The image quality was mildly variable on the 
night the data were recorded, and exposures in 
which the image quality was noticeably poorer 
than average were not used. A total of 12 exposures were combined to 
construct the final image. Stars in the final dataset have FWHM = 0.10 arcsec.

	The deep $K-$band image discussed by Davidge et al. (2000), which 
was recorded with the Hokupa'a AO system and QUIRC imager on GN during 
July 2000, was also used in this study. The image was rotated and re-sampled 
to match the orientation and pixel scale of the ALTAIR$+$NIRI dataset. 
The FWHM of these data are comparable to those 
obtained with ALTAIR, and additional details of these 
observations are discussed by Davidge et al. (2000). 

\section{RESULTS}

	The primary statistic used in this study is $\Delta K = 
K_{Hokupa'a} - K_{ALTAIR}$, which is a measure of photometric stability and the 
amplitude of light variations among variable stars. This quantity was computed by 
taking the difference between the brightnesses of individual stars in the ALTAIR 
and Hokupa'a images, as measured with the point spread function (PSF)-fitting 
routine ALLSTAR (Stetson \& Harris 1988). The $(K_{ALTAIR}, \Delta K)$ diagrams 
for stars in two radial intervals, the inner and outer boundaries of which were 
selected (1) to avoid the very crowded inner few arcsec of the galaxy, and (2) 
to balance roughly the number of stars in each annulus near the bright end, 
are shown in Figure 1.

	$\Delta K$ ranges roughly from --1 to $+1$ 
near the bright end of Figure 1, with the majority of points 
concentrated between $\Delta K = -0.3$ and $+0.3$. The histogram distribution 
of $\Delta K$ for stars in each annulus with $K_{ALTAIR}$ between 16.3 and 
16.8 is shown in Figure 2. This particular brightness interval was 
selected to sample (1) a moderately large number of stars at a brightness where 
incompleteness is not significant (see below), and (2) 
stars that are well above the RGB-tip. Based on the calibration derived by Ferraro 
et al. (2000) from globular clusters and assuming an old solar-metallicity 
population, then the RGB-tip occurs near $K = 17.4$ in M32 if the 
distance modulus is $\mu_0 = 24.4$ (van den Bergh 2000). 

	The $\Delta K$ distributions of stars in the two annuli have a similar 
gaussian-like appearance, with a standard deviation $\sigma = \pm 0.3$ mag. 
At least part of the spread in $\Delta K$ is a consequence 
of random photometric errors. Artificial star experiments were run to assess the 
size of the random uncertainties in these data, and these experiments predict that 
the dispersion due to random errors is $\pm 0.13$ mag in each interval, which is 
sigificantly smaller than the observed dispersion in $\Delta K$. The artificial 
star experiments also indicate that the data in both radial intervals are $\sim 
80\%$ complete when $K$ is between 16.3 and 16.8. 

	That the $\Delta K$ distributions near the center of M32 are much 
broader than expected from observational errors indicates that many of 
the stars are photometrically variable. In fact, the $\Delta K$ distributions 
measured in M32 are remarkably consistent with what is expected from 
a pure population of long period variables (LPVs) like those in the Galactic 
bulge. To demonstrate this point, the extensive observations of LPVs in the 
Sgr I field of the Galactic bulge obtained by Glass et al. (1995) were used to 
create a reference $\Delta K$ distribution for comparison with the M32 data. 
The Glass et al. (1995) observations are ideal for this purpose because they 
sample a moderately large number of stars with intrinsic brightnesses that are 
comparable to the target objects in M32, while also spanning a time baseline 
that exceeds that between the Hokupa'a and ALTAIR observations.

	The observations for each star in Table 1 of Glass et al. (1995) were 
paired, with the restriction that the difference in epochs between 
the observations in each pair was at least 
1000 days in order to match the approximate time difference between the Hokupa'a 
and ALTAIR datasets. The difference in $K-$band brightness for each data pair, 
which corresponds to $\Delta K$, was then computed. Typically 4 -- 5 independent 
(i.e. such that each observation was used only in one pair) $\Delta K$ values 
were computed for each star. Large amplitude IRAS variables were 
not included, as these objects have very red colors, and there is no evidence 
for such a population in M32 (Davidge 2000; Davidge et al. 2000).

	The histogram distribution of the $\Delta K$ measurements constructed from 
the Glass et al. (1995) observations was convolved with a $\sigma = 0.13$ 
gaussian to account for the smearing introduced by random uncertainties in the M32 
dataset and the result, scaled to match the number of stars in each radial M32 
interval, is compared with the M32 $\Delta K$ distributions in Figure 
2. The Galactic and M32 $\Delta K$ distributions agree within the estimated 
$2-\sigma$ uncertainties in the vast majority of bins, and a Kolmogorov-Smirnov 
test indicates that the Galactic and M32 $\Delta K$ distributions are not 
significantly different.
 
	It is likely that some of the bright stars in M32 are not 
variable, and so composite models consisting of a non-variable 
component, the $\Delta K$ distribution of which was 
represented by a Gaussian with $\sigma = 0.13$ mag, and Galactic bulge LPVs were 
also considered. Model $\Delta K$ distributions that include a 
minority non-variable component give a better match to the M32 $\Delta K$ 
distribution than the LPV--only model. The best agreement 
with the M32 $\Delta K$ distribution occurs when $80\%$ of the stars are 
LPVs, and this model is compared with the M32 $\Delta K$ distributions in 
Figure 3. The M32 and model $\Delta K$ distributions differ at the 
$95\%$ confidence level or greater when less than $60\%$ of the stars are LPVs. 
Thus, the $\Delta K$ distribution of the brightest AGB stars near the center 
of M32 is consistent with at least $60\%$ of these objects having an amplitude 
distribution similar to LPVs in the Galactic bulge.

\section{DISCUSSION \& SUMMARY}

	We have measured the change in $K-$band magnitude of bright AGB stars 
near the center of M32 over a 3 year baseline. The distribution of brightness 
differences, which is a measure of the amplitude of light 
variations, has been compared with that expected for
LPVs detected at visible-red wavelengths in the Galactic bulge. It is concluded 
that at least 60\% of the bright AGB stars near the center of M32 are LPVs with 
amplitudes similar to LPVs in the Galactic bulge.

	Rejkuba et al. (2003) discuss the properties of LPVs in the 
nearby elliptical galaxy NGC 5128 (Cen A). They find that (1) the majority 
of LPVs are like those in the Galactic bulge, and (2) in one of the 
fields they studied at least 70\% of the AGB stars are LPVs. These results are 
very similar to what we find in M32. However, Rejkuba et al. (2003) 
also found that 10\% of the LPVs had periods that were comparable to the 
IRAS variables detected in the Galactic bulge, which presumably come from a 
population that is younger than the shorter period LPVs. Very long period, 
large amplitude LPVs were intentionally excluded from the comparison 
in \S 3. The distribution of $\Delta K$ values 
computed from the IRAS variables studied by Glass et al. (1995) is very 
different from that computed from the variables discovered at visible-red 
wavelengths, and the standard deviation in the $\Delta K$ distribution 
computed solely from IRAS variables is $\sigma = \pm 1.0$ mag, which is almost 
three times that measured for LPVs detected at visible-red wavelengths. There are 
some bright objects in the $(K, \Delta K)$ diagram that have $\Delta K = \pm 1$, 
but these are probably the extreme members of a lower amplitude LPV distribution. 
There is a selection effect working against the detection of large amplitude 
variables in the current study, as they may fall below the detection limit when at 
the faint points in their light curves; however, these variables 
should be easy to detect when near the peak of their light curves. While we can 
not discount the presence of a modest number of large amplitude variables in M32 
with these data, it is worth noting that these objects tend to have very 
red colors, and the narrow, well-defined AGB sequence 
see near the center of M32 (Davidge et al. 2000) and at moderately large radii 
(Davidge 2000) argues against the presence of such a population.

	Surveys of variable stars in Baade's Window (BW) reveal objects with a 
range of characteristics that match those seen 
in the Magellanic Clouds (Glass \& Schultheis 2003). Some of the variables in 
BW have modest amplitudes, and such a population is likely to be present in M32; 
indeed, some fraction of the `non-variable' component included in the models 
may actually be small amplitude variables. However, the number of small 
amplitude variables in our sample is likely modest. 
Assuming that the distance modulus of M32 is $\mu_0 = 24.4$ (van den Bergh 
2000), and that the distance modulus of the Sgr I field is $\mu_0 = 14.7$ (Glass 
et al. 1995), then the range of brightnesses used to construct the M32 $\Delta 
K$ distribution in Figure 2 corresponds to K$_0$ between 6.6 and 7.1 in BW. It is 
evident from Figure 7 of Glass \& Schultheis (2003) that the variable stars in BW 
in this brightness interval tend to have log(P) $> 2$, and hence are larger 
amplitude variables. Lower amplitude variables occur in large numbers in BW only 
when $K_0 > 7.1$, and it can be anticipated that these objects 
will become more significant in M32 at fainter brightnesses than considered here.

	We close by noting that with the up-coming generation of AO-equipped 
$30+$ metre telescopes it will be possible to resolve bright stars in spheroids 
in the Virgo cluster and beyond (e.g. Anthony et al. 2003). A traditional 
characterization of LPVs in galaxies at this distance, involving observations 
over a large number of nights to obtain light curves, may only be practical 
for a modest number of systems because of the long integration times 
involved and time assignment pressures imposed by the need for nights 
with superb imaging conditions. The technique employed here, which 
requires observations covering only two epochs, provides 
an economical means of surveying the amplitude characteristics of LPVs 
in a large number of systems.

\clearpage

\begin{center}
FIGURE CAPTION
\end{center}

\figcaption
[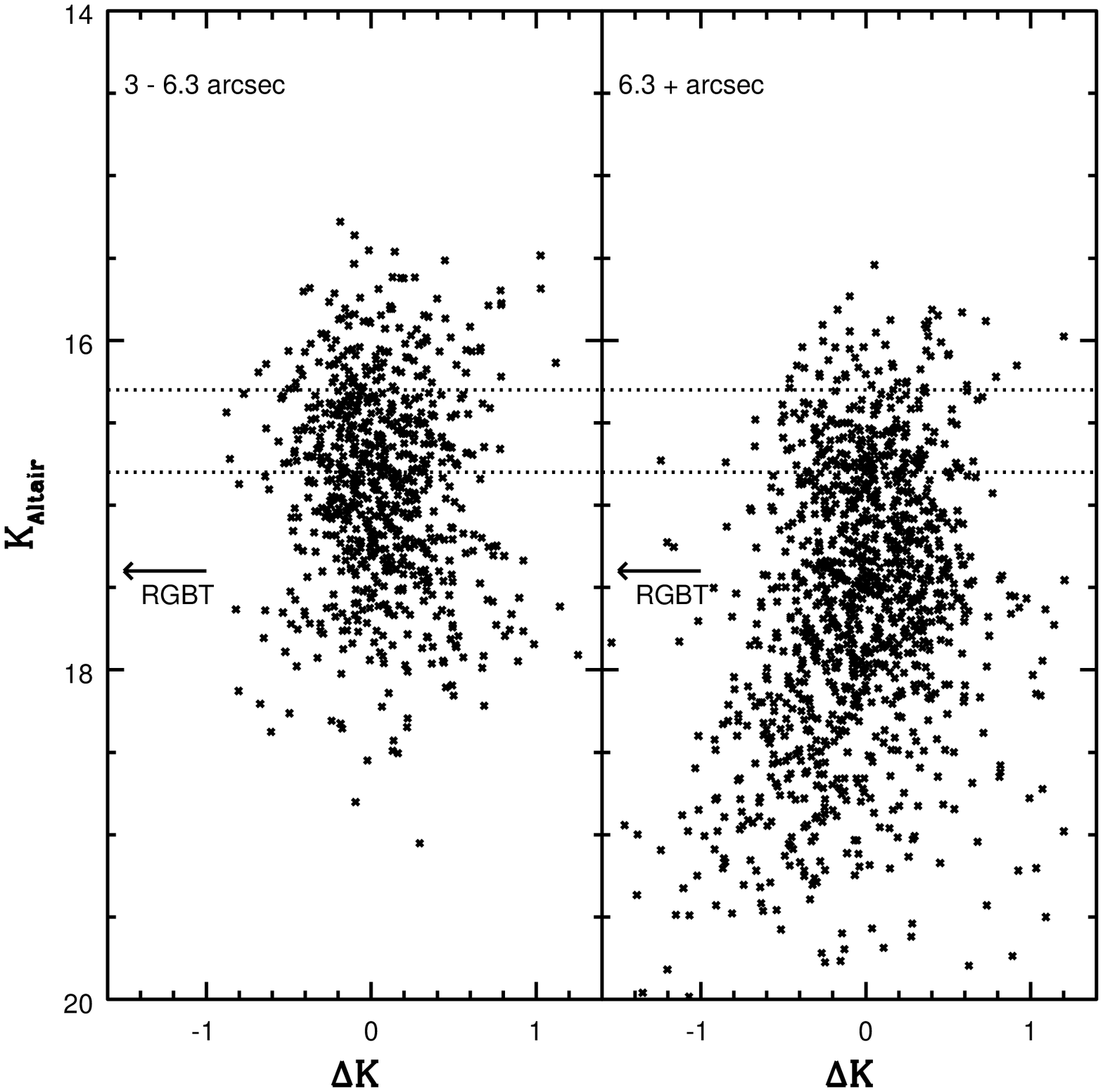]
{A comparison between the $K-$band brightnesses of stars 
near the center of M32, as measured from the Hokupa'a and ALTAIR datasets. 
$\Delta K$ is the difference between the Hokupa'a and ALTAIR brightnesses. 
The dashed lines mark the upper and lower boundaries used to compute the 
$\Delta K$ distribution in Figure 2. The brightness of the RGB-tip in 
an old solar-metallicity population, computed using the globular cluster-based 
Ferraro et al. (2000) calibration, and assuming $\mu_0 = 24.4$ for 
M32 (van den Bergh 2000), is also indicated.} 

\figcaption
[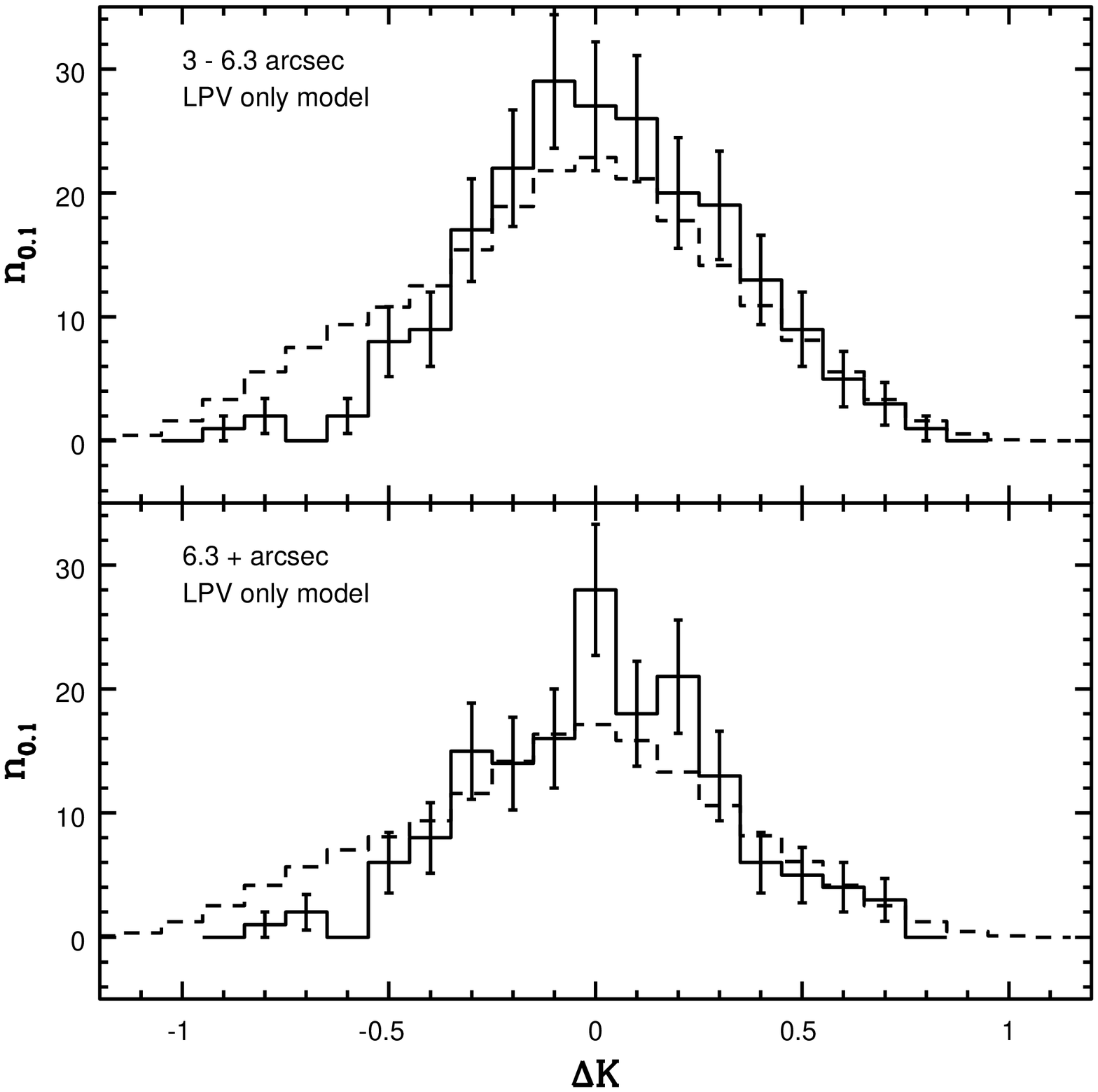]
{The histogram distribution of $\Delta K$ in two radial intervals in M32. 
n$_{0.1}$ is the number of stars with $K$ between 16.3 and 16.8 per 0.1 mag 
interval in $\Delta K$. The errorbars show $1-\sigma$ uncertainties due to 
Poisson statistics. The dashed line is the $\Delta K$ distribution of LPVs in the 
Galactic bulge, as generated from photometric measurements in Table 1 of Glass et 
al. (1995) using the procedure described in the text. 
The Galactic LPV distribution in each panel has been 
convolved with a Gaussian distribution with $\sigma = \pm 0.13$ to account for 
the random uncertainies in the M32 observations, as determined 
from artificial star experiments, and then scaled to match 
the total number of stars with $K$ between 16.3 and 16.8 in each M32 region.}

\figcaption
[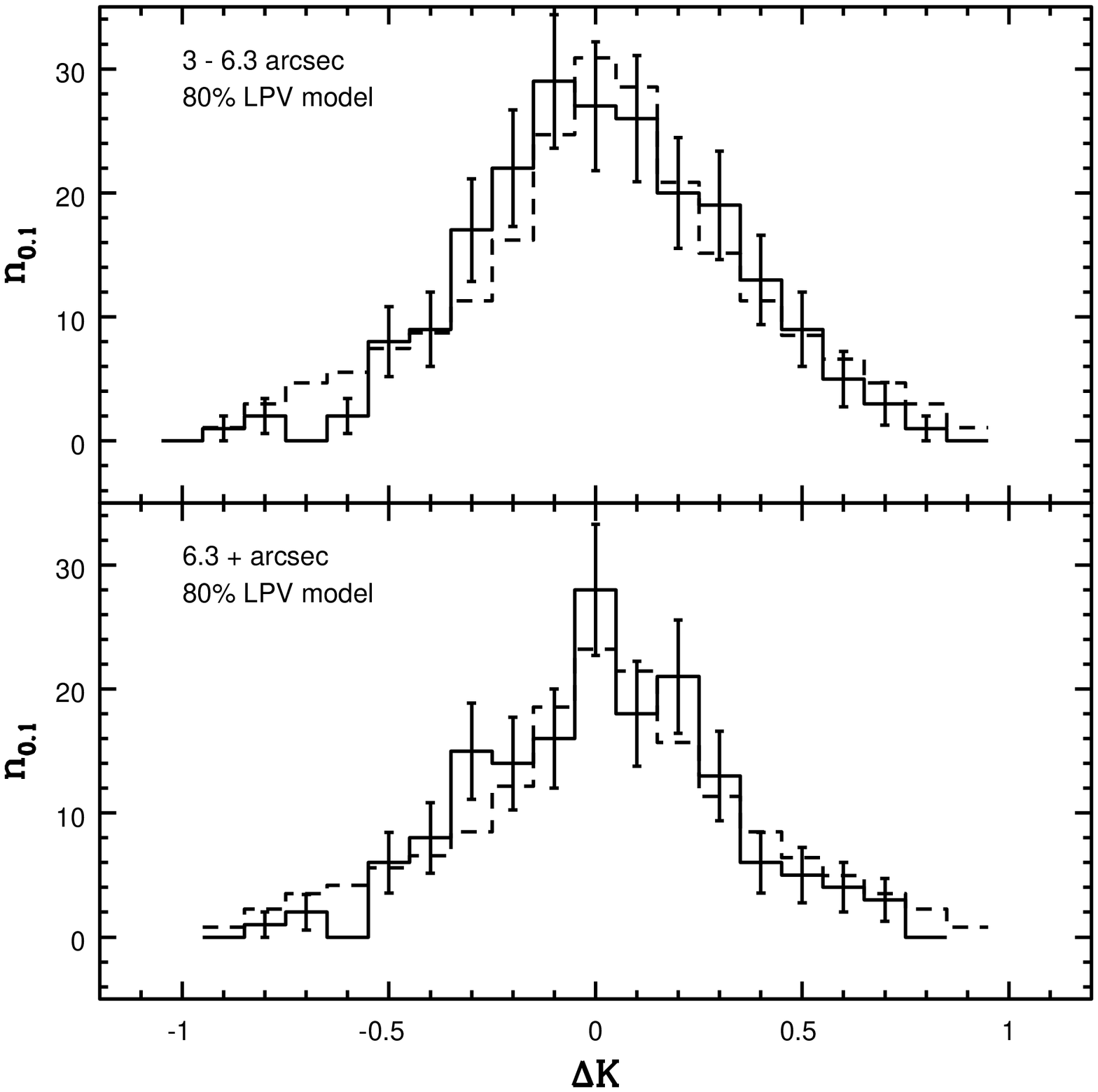]
{The same as Figure 2, but with the dashed line showing the model $\Delta K$ 
distribution if 
$80\%$ of the stars are Galactic bulge LPVs. Note that the agreement between 
the M32 and model $\Delta K$ distributions is much better than in Figure 2.}

\end{document}